# Defect propagation in one-, two-, and three-dimensional compounds doped by magnetic atoms

A. Furrer[1,*], A. Podlesnyak[2], K.W. Krämer[3], and Th. Strässle[1]

[1]Laboratory for Neutron Scattering, Paul Scherrer Institut, CH-5232 Villigen PSI, Switzerland

[2]Quantum Condensed Matter Division, Oak Ridge National Laboratory, Oak Ridge, TN 37831-6473, USA

[3]Department of Chemistry and Biochemistry, University of Bern, CH-3012 Bern, Switzerland

Inelastic neutron scattering experiments were performed to study manganese(II) dimer excitations in the diluted one-, two-, and three-dimensional compounds $CsMn_xMg_{1-x}Br_3$, $K_2Mn_xZn_{1-x}F_4$, and $KMn_xZn_{1-x}F_3$ ($x \leq 0.10$), respectively. The transitions from the ground-state singlet to the excited triplet, split into a doublet and a singlet due to the single-ion anisotropy, exhibit remarkable fine structures. These unusual features are attributed to local structural inhomogeneities induced by the dopant Mn atoms which act like lattice defects. Statistical models support the theoretically predicted decay of atomic displacements according to $1/r^2$, $1/r$, and constant (for three-, two-, and one-dimensional compounds, respectively) where $r$ denotes the distance of the displaced atoms from the defect. The observed fine structures allow a direct determination of the local exchange interactions $J$, and the local intradimer distances $R$ can be derived through the linear law $dJ/dR$.

[*]Corresponding author: albert.furrer@psi.ch

# I. INTRODUCTION

In two previous publications we reported on the propagation of defects in doped magnetic materials based on data from inelastic neutron scattering (INS) experiments performed for the one-dimensional compound $CsMn_xMg_{1-x}Br_3$ (x=0.05, 0.10) [1] as well as for the two- and three-dimensional compounds $K_2Mn_{0.10}Zn_{0.90}F_4$ and $KMn_{0.10}Zn_{0.90}F_3$, respectively [2]. The partial substitution of nonmagnetic $Mg^{2+}$ or $Zn^{2+}$ ions by $Mn^{2+}$ ions results in the creation of Mn multimers as shown for the case of Mn dimers in Fig. 1. The low Mn concentration x≤0.10 maximizes the statistical creation of dimers with respect to trimers, tetramers, etc., and minimizes the effect of next-nearest-neighbor interactions as well. The spin Hamiltonian of Mn dimers is given by

$$H = -2J\mathbf{s_1} \cdot \mathbf{s_2} - D\left[\left(s_1^z\right)^2 + \left(s_2^z\right)^2\right] \qquad (1)$$

where $\mathbf{s_i}$ denotes the spin operator of the Mn ions (with $s_i$=5/2), $J$ the bilinear exchange parameter, and $D$ the axial single-ion anisotropy parameter. Eq. (1) gives rise to a characteristic excitation spectrum, and transitions between different dimer states $|S,M>$ can be directly measured by INS experiments, where $S$ is the total dimer spin and $-S \leq M \leq S$. The resulting energy level scheme is shown in Fig. 2. In Refs. [1,2] we focused on the transition from the ground-state singlet $|0,0>$ to the excited triplet state, which is composed of a doublet $|1,\pm1>$ and a singlet $|1,0>$ split by the single-ion anisotropy. The observed transitions exhibit remarkable fine structures which are attributed to local structural inhomogeneities around the Mn dimers. More specifically, the substitution of Mg or Zn ions by Mn ions produces an internal chemical pressure due to the differences of the ionic radii ($r_{Mn}$=0.83 Å > $r_{Zn}$=0.74 Å > $r_{Mg}$=0.72 Å for divalent ions with six-fold coordination [3]), so that the atomic positions have to rearrange in the vicinity of the dopant Mn ions, which behave like defects. The defect problem was thoroughly investigated by Krivoglaz [4] who showed that the atomic displacements in three-dimensional crystals decay asymptotically as $1/r^2$ ($r$ denotes the distance of the displaced atoms from the defect). For defects in two-dimensional crystals the atomic displacements are expected to decrease according to $1/r$, and in one-dimensional crystals the atomic displacements do not decrease with increasing distance $r$ at all. These predictions were qualitatively confirmed by statistical models in Refs. [1,2].

The question arises whether the hypothesis of local structural distortions postulated in Refs. [1,2] is both necessary and sufficient to describe the observed fine structures. The agreement between the experimental findings and the statistical model calculations may be accidental, so that it is highly desirable to consider additional criteria. Further hypothesis tests include, *e.g.*, (i) the variation of parameters relevant for the model, (ii) the verification of the model assumptions, (iii) the extension of the experimental data to hitherto unexplored properties, and (iv) the statistical improvement of the existing data. We therefore performed further INS investigations along these lines which provided additional important results and thereby strengthened the underlying hypotheses. More specifically, we extended the experiments for the one-dimensional compound $CsMn_{0.10}Mg_{0.90}Br_3$ to include not only the lowest-lying $|0,0>\rightarrow|1,\pm1>$ transition A, but also the higher-lying $|0,0>\rightarrow|1,0>$ transition B (see Fig. 2). The observation of almost identical fine structures for two dimer transitions with different wave functions unambiguously supports our analysis in terms of atomic displacements induced by defects and excludes other possible interpretations (*e.g.*, in terms of next-nearest-neighbor exchange interactions). Moreover, the new data benefit from an improved statistics (three times better than the data of Ref. 1). For the two- and three-dimensional compounds $K_2Mn_xZn_{1-x}F_4$ and $KMn_xZn_{1-x}F_3$, respectively, the single-ion anisotropy was neglected in Ref. 2. However, new experiments on very dilute compounds showed that the single-ion anisotropy splits the excited triplet in a substantial manner, which has to be considered in the interpretation of the data. In addition, we were able to observe fine structures for different Mn concentrations *x*, which is an important parameter of the applied statistical model.

**II. EXPERIMENTAL**

The synthesis and the characterization of the samples were described in Refs. [1,2]. The INS experiments were performed with use of the time-of-flight spectrometer CNCS [5] at the spallation neutron source SNS at Oak Ridge National Laboratory. The samples were enclosed in Al cylinders (12 mm diameter, 45 mm height) and placed into a He cryostat to achieve temperatures T>2 K. Incoming neutrons with energies of 1.55 meV and 2.50 meV were used, giving rise to unprecedented energy resolutions of Gaussian shape with full width at half maximum (FWHM)=11 µeV at $E \approx 0.8$ meV and FWHM=17 µeV (15 µeV) at $E \approx 1.77$ meV (1.91 meV), respectively, where *E* denotes the energy transfer.

**III. THEORETICAL BACKGROUND**

Fig. 1 visualizes the atomic positions around isolated Mn dimers in one-, two-, and three-dimensional structures investigated in the present work. For x<<1 these positions are mainly occupied by Mg or Zn atoms, which are occasionally replaced by Mn atoms according to statistics. In the following we summarize the corresponding statistical laws compatible with the theoretical predictions of Krivoglaz [4].

For the one-dimensional case as sketched for a $Mn_xMg_{1-x}$ chain in Fig. 1(a), the probabilities $p_m(x)$ for having *m* Mn atoms on both sides of the central Mn dimer in a chain of length 2*n* are given by [1,6]

$$p_0(x) = (1-x)^{2n}$$
$$p_1(x) = 2\binom{n}{1}x(1-x)^{2n-1}$$
$$p_2(x) = \left[2\binom{n}{2} + \binom{n}{1}\binom{n}{1}\right]x^2(1-x)^{2n-2} \qquad (2)$$
$$p_3(x) = 2\left[\binom{n}{3} + \binom{n}{2}\binom{n}{1}\right]x^3(1-x)^{2n-3}$$

etc.

The chain length $2n$ has to be chosen such that the sum rule $\Sigma_m p_m(x)=1$ and the condition $2n \geq m$ are satisfied. These criteria are fulfilled for $n=1/x$. According to Krivoglaz [4] the decay of atomic displacements is governed by the number $m$ of Mn atoms and not by their specific arrangement in the chain.

For the three-dimensional case as sketched in Fig. 1(c) for mixed Mn/Zn structures, the probability for having $m$ Mn ions at the $n$ nearest-neighbor positions is given by [2]

$$p_m(x) = \binom{n}{m}x^m(1-x)^{n-m} \qquad . \qquad (3)$$

Because of the short-range $1/r^2$ decay law predicted by Krivoglaz [4] we do not consider further-distant-neighbor positions in Eq. (3).

For two-dimensional systems the $1/r$ decay law has a long-range nature, so that the nearest-neighbor approximation is not valid as proven in Ref. 2. The statistical model was extended to include also the next-nearest-neighbor positions as sketched in Fig. 1(b), which turned out to provide a reasonable agreement with the experimental data. Eq. (3) applies for the calculation of the corresponding probabilities $p_m(x)$.

Eqs. (2) and (3) will be used to calculate the intensities of the fine structures observed in the excitation spectra. The positions of the fine-structure lines (indexed by $m=0,1,2,...$) will be analyzed by the spin Hamiltonian (1) in order to determine the local exchange parameters $J_m$. Through the linear law $dJ/dR$ the local intradimer distances $R_m$ can then be derived as a function of internal pressure induced by the substitutional Mn atoms.

## IV. RESULTS AND DATA ANALYSIS

### A. The one-dimensional compound CsMn$_{0.10}$Mg$_{0.90}$Br$_3$

The compound CsMn$_x$Mg$_{1-x}$Br$_3$ crystallizes in the hexagonal space group $P6_3/mmc$ with chains of face-sharing MBr$_6$ (M=Mn, Mg) octahedra along the $c$ axis. Energy spectra of neutrons scattered from CsMn$_{0.10}$Mg$_{0.90}$Br$_3$ at T=2 K are shown in Figs. 3(a) and 3(b), which correspond to the ground-state Mn dimer transitions $|0,0\rangle \rightarrow |1,\pm1\rangle$ and $|0,0\rangle \rightarrow |1,0\rangle$, respectively, with an intensity ratio of 3:1 in agreement with the neutron cross-section for dimer transitions [7] and moduli of the scattering vector $0.6 \leq Q \leq 1.4$ Å$^{-1}$. Both spectra exhibit very similar fine structures which are well described by a superposition of five individual lines ($m=0,...,4$) of Gaussian shape with almost equidistant energy spacings. The adjustable parameters in the least-squares fitting procedure were a linear background, the positions and the amplitudes of the individual lines, whereas the linewidth was fixed at the instrumental energy resolutions of 17 µeV (a) and 15 µeV (b). The least-squares fitting procedure rapidly converged to the results displayed as dashed lines in Fig. 3, and the line positions are listed in Table I.

Although the Mn$_x$Mg$_{1-x}$ chains sketched in Fig. 1(a) are not strictly elastically decoupled from the neighboring chains, they exhibit sufficient one-dimensional character to treat the structural distortions around the central Mn dimer by the statistical model described by Eq. (2) which was applied to calculate the intensities of the individual lines. The resulting probabilities $p_m(x)$ for $0 \leq m \leq 4$, scaled to the amplitudes of the individual lines, are indicated as vertical bars in Fig. 3. The agreement between the amplitudes of the individual lines and the calculated probabilities $p_m(x)$ is remarkably good with standard deviations $\chi^2=1.2$ (a) and $\chi^2=1.8$ (b).

Table I lists the local exchange interactions $J_m$ derived from Eq. (1). With increasing number $m$ the values of $J_m$ are continuously inreasing, whereas the anisotropy parameter $D=0.0207(3)$ meV turns out to be independent of $m$. The increasing values of $J_m$ result from a compression of the local intradimer Mn-Mn distances $R_m$ due to the internal chemical pressure exerted by the dopant Mn ions in the chain. Applying the relation $dJ/dR=3.6(3)$ meV/Å [8] yields the $R_m$ values listed in Table I.

The data analysis carried out in Ref. [1] was based on a spin Hamiltonian in which Eq. (1) included a biquadratic term $K(\mathbf{s_1} \cdot \mathbf{s_2})^2$ (with K=0.0086 meV). As a consequence, for $D=0$ the singlet-triplet splitting $|S=0\rangle \rightarrow |S=1\rangle$ is modified from $2J$ (see Fig. 2) to $2J+16.5K$ [7], which explains the rather large differences of the resulting $J_m$ values. On the other hand, the resulting $R_m$ values are practically unaffected by the neglect of the biquadratic term.

## B. The two-dimensional compound $K_2Mn_xZn_{1-x}F_4$

The compound $K_2Mn_xZn_{1-x}F_4$ crystallizes in the tetragonal space group $I4/mmm$. As outlined in Ref. 2, the elastic energy along the $z$ direction is considerably weaker than in the $x$-$y$ plane, thus it is well justified to describe the elastic properties by a two-dimensional model as sketched in Fig. 1(b). Fig. 4 shows energy spectra of neutrons scattered from $K_2Mn_xZn_{1-x}F_4$ at T=2 K with $x$=0.02 (a) and $x$=0.10 (b). The peaks A and B in Fig. 4(a) correspond to the Mn dimer ground-state transitions shown in Fig. 2. The spectrum was least-squares fitted by two Gaussian lines with FWHM=21(1) µeV. The intensity ratio $I_A/I_B$=2.8(3) is in agreement with the calculated ratio of 3.0 for moduli of the scattering vector 0.5≤Q≤1.4 Å$^{-1}$. The analysis based on Eq. (1) yields $J$=-0.4205(5) meV and $D$=0.0052(2) meV, and the anisotropy-induced splitting between the states |1,±1> and |1,0> amounts to 33(1) µeV.

Fig. 4(b) shows the data for $x$=0.10 and their subdivision into seven individual lines of Gaussian shape with FWHM=18(2) µeV and almost equidistant energy spacings of 16 µeV taken from Ref. 2. The overall spectrum is shifted to lower energies compared to the $x$=0.02 data due to a slight expansion of the lattice parameters. The local structural model applied in Ref. 2 was based on the assumption of a negligibly small anisotropy splitting being smaller than the width of the individual lines, which is certainly not true as proven by Fig. 4(a). In fact, the anisotropy splitting (33 µeV) is almost exactly twice as large as the energy spacings between the individual lines (16 µeV). This has to be considered in the modelling based on Eq. (3), *i.e.*, the individual lines $m$ result from a 3:1 weighted superposition of fine structures associated with both peaks A and B (the latter shifted upwards by the anisotropy splitting of 33 µeV). This procedure is well justified, since the data presented in Sec. IV.A gave evidence for almost identical fine structures for both Mn dimer transitions |0,0>→|1,±1> (corresponding to peak A) and |0,0>→|1,0> (corresponding to peak B). The resulting probabilities calculated from Eq. (3) with $m$=24, *i.e.*, including the ten nearest neighbours and the 14 next-nearest neighbors as sketched in Fig. 1(b), are indicated as vertical bars in Fig. 4(b). The agreement between the amplitudes of the individual lines and the weighted sum of the calculated probabilities $p_{mA}(x)$ and $p_{mB}(x)$ is reasonably good with a standard deviation $\chi^2$=2.8. Compared to the analysis presented in Ref. 2, the center of gravity of the calculated amplitudes is slightly shifted to higher energies due to the inclusion of the anisotropy splitting.

Table II lists the local exchange interactions $J_m$ derived from Eq. (1) with $D$=0.0052 meV determined for the x=0.02 data as well as the local intradimer Mn-Mn distances $R_m$. The latter were calculated from the relation $dJ/dR$=2.6(4) meV/Å [2]. The present data slightly differ from our earlier results ($\Delta J_m$≈0.005 meV, $\Delta R_m$≈0.002 Å) due to the disregard of the single-ion anisotropy in Ref. [2].

## C. The three-dimensional compound KMn$_x$Zn$_{1-x}$F$_3$

KMn$_x$Zn$_{1-x}$F$_3$ crystallizes in the cubic space group *Pm3m*, *i.e.*, it is a truly three-dimensional compound as sketched in Fig. 1(c). Fig. 5 shows energy spectra of neutrons scattered from KMn$_x$Zn$_{1-x}$F$_3$ at T=2 K with *x*=0.01 (a) and *x*=0.10 (b). The spectrum of Fig. 5(a) was least-squares fitted by three Gaussian lines with FWHM=17(1) µeV. We attribute the peaks A and B to the Mn dimer ground-state transitions shown in Fig. 2. The analysis based on Eq. (1) yields *J*=-0.4100(4) meV and *D*=0.0053(2) meV, and the anisotropy-induced splitting between the states |1,±1> and |1,0> amounts to 33(1) µeV. The intensity ratio $I_A/I_B$=2.4(2), however, is considerably smaller than the calculated ratio of 3.0 for moduli of the scattering vector 0.5≤Q≤1.4 Å$^{-1}$. This discrepancy as well as the presence of the tiny peak C are due to local structural inhomogeneities as discussed below.

Fig. 5(b) shows the data for *x*=0.10 and their subdivision into seven individual lines of Gaussian shape with FWHM=27 µeV and almost equidistant energy spacings of 30 µeV taken from Ref. 2. The overall spectrum is shifted to lower energies compared to the *x*=0.01 data due to a slight expansion of the lattice parameters. The local structural model applied in Ref. 2 was based on the assumption of a negligibly small anisotropy splitting being smaller than the width of the individual lines, which is in contradiction to Fig. 5(a). In fact, the anisotropy splitting (33 µeV) is roughly equal to the energy spacings between the individual lines (30 µeV). This has to be considered in the model calculations as discussed in Sec. IV.B. The resulting probabilities calculated from Eq. (3) with $m$=26 nearest neighbours as sketched in Fig. 1(c) are indicated as vertical bars in Fig. 5. The agreement between the amplitudes of the individual lines and the 3:1 weighted sum of the calculated probabilities $p_{mA}(x)$ and $p_{mB}(x)$ is reasonably good with standard deviations $\chi^2$=3.8 (a) and 2.7 (b). Compared to the analysis presented in Ref. 2, the center of gravity of the calculated amplitudes in Fig. 5(b) is slightly shifted to higher energies due to the inclusion of the anisotropy splitting.

Table II lists the local exchange interactions $J_m$ derived from Eq. (1) with $D$=0.0053 meV determined for the x=0.01 data as well as the local intradimer Mn-Mn distances $R_m$. The latter were calculated from the relation $dJ/dR$=3.3(6) meV/Å [2]. The present data slightly differ from our earlier results ($\Delta J_m \approx 0.005$ meV, $\Delta R_m \approx 0.002$ Å) due to the disregard of the single-ion anisotropy in Ref. [2].

## V. CONCLUDING REMARKS

Our new INS data of Mn dimer excitations in $KMn_xZn_{1-x}F_3$, $K_2Mn_xZn_{1-x}F_4$, and $CsMn_xMg_{1-x}Br_3$ and their fine-structure analyses support the theoretically predicted decay of atomic displacements according to $1/r^2$, $1/r$, and constant [4] resulting from substitutional defects (Zn or Mg atoms replaced by Mn atoms) in three-, two-, and one-dimensional crystals, respectively. In particular, the extension of the data to different Mn concentrations $x$ and to different Mn dimer transitions, both being crucial inputs to the statistical models, as well as a correct treatment of the anisotropy splitting unambiguously confirm the hypothesis of local structural inhomogeneities being the origin of the observed fine structures put forward in Refs. [1,2].

For the one-dimensional compound, the statistical model defined by Eq. (2) was rigorously applied without any approximations. As shown in Fig. 3, the agreement with the experimental data is excellent and can claim quantitative accuracy with the theoretical prediction of a constant decay of atomic displacements [4]. The fact that the observed fine-structure lines are limited by the instrumental energy resolution gives further support for the underlying model. More specifically, the defects statistically present in the chains clearly do not produce a continuous local displacement pattern, but the displacements lock in at discrete values depending on the number of defects, irrespective of their positions in the chain.

On the other hand, for the three- and two-dimensional compounds some approximations were used to calculate the decay laws $1/r^2$ and $1/r$, respectively [4]. In particular, we included in Eq. (3) only the nearest-neighbor positions for the $1/r^2$ law and in addition the next-nearest-neighbor positions for the $1/r$ law which, however, is not completely sufficient to model the asymptotic behavior of the decay of atomic displacements. Nevertheless, we found reasonable agreement with the experimental data as shown in Figs. 4 and 5, thus our analyses support the theoretical predictions at least on a qualitative level. Some caution is advised in view of the widths of the fine-structure lines being larger than the instrumental energy resolution. The increased linewidths are partly due to the specific treatment of the anisotropy splitting explained in Secs. IV.B and IV.C, but the major enhancement certainly results from the above mentioned shortcomings of the model approximations. Possible improvements of the model include the consideration of further-distant-neighbor positions as well as a distinction of the neighbor positions with respect to the dimer axis. However, all these model extensions result in many additional fine-structure lines, whose detection would require INS experiments with considerably higher energy resolution than currently available at first-class neutron sources.

In conclusion, we emphasize that the present work does not only support the theoretically predicted decay laws for defect propagation, but the applied experimental method also provides information on both local exchange interactions and local structural effects. Modern spectroscopies measure exchange couplings with a precision of *dJ/J*<0.01, thus spatial resolutions of *dR*<0.01 Å can be achieved on an absolute scale and typically ten times better on a relative scale (see Tables I and II). In addition, we have shown that the exchange interaction in doped materials is no longer uniformly distributed, but it can deviate up to 10% from the average exchange (see Table II). These effects appear to be of increasing interest and importance in investigations of doped superconductors and quantum spin systems [9,10].

## ACKNOWLEDMENT

The assistance of D. Biner (University of Bern) in the synthesis of the samples is gratefully acknowledged. Research at Oak Ridge National Laboratory's Spallation Neutron Source was supported by the Scientific User Facilities Division, Office of Basic Energy Sciences, US Department of Energy.

TABLE I. Analysis of experimental data observed for $CsMn_{0.10}Mg_{0.90}Br_3$. $E_a$ and $E_b$ denote the energy transfers of the individual lines $m$ displayed in Figs. 1(a) and 1(b), respectively. The exchange couplings $J_m$ and the local Mn-Mn distances $R_m$ associated with each pair of lines were derived as explained in the text. Relative error bars are given for $J_m$ and $R_m$.

| $m$ | $E_a$ [meV] | $E_b$ [meV] | $J_m$ [meV] | $R_m$ [Å] |
|---|---|---|---|---|
| 0 | 1.742(3) | 1.877(5) | -0.8925(8) | 3.2296(3) |
| 1 | 1.760(2) | 1.893(3) | -0.9014(6) | 3.2272(2) |
| 2 | 1.774(2) | 1.906(3) | -0.9088(6) | 3.2251(2) |
| 3 | 1.788(2) | 1.919(3) | -0.9153(6) | 3.2233(2) |
| 4 | 1.803(3) | 1.934(5) | -0.9228(9) | 3.2212(3) |

TABLE II. Analysis of experimental data observed for $K_2Mn_{0.10}Zn_{0.90}F_4$ and $KMn_{0.10}Zn_{0.90}F_3$. $E_m$ denotes the energy transfer of the individual lines marked by $m_A$ as displayed in Figs. 3(b) and 4(b). The exchange couplings $J_m$ and the local Mn-Mn distances $R_m$ associated with each line $m_A$ were derived as explained in the text. Relative error bars are given for $J_m$ and $R_m$.

| $m_A$ | $K_2Mn_{0.10}Zn_{0.90}F_4$ | | | $KMn_{0.10}Zn_{0.90}F_3$ | | |
|---|---|---|---|---|---|---|
|  | $E_m$ [meV] | $J_m$ [meV] | $R_m$ [Å] | $E_m$ [meV] | $J_m$ [meV] | $R_m$ [Å] |
| 0 | 0.780(3) | -0.3953(32) | 4.0546(7) | 0.729(4) | -0.3700(31) | 4.0674(7) |
| 1 | 0.797(2) | -0.4039(21) | 4.0513(5) | 0.764(3) | -0.3873(22) | 4.0621(5) |
| 2 | 0.812(1) | -0.4115(14) | 4.0484(4) | 0.793(1) | -0.4020(12) | 4.0577(3) |
| 3 | 0.827(2) | -0.4191(24) | 4.0455(5) | 0.822(2) | -0.4167(18) | 4.0532(4) |
| 4 | 0.846(2) | -0.4282(19) | 4.0420(4) | 0.850(3) | -0.4304(19) | 4.0491(4) |
| 5 | 0.861(3) | -0.4358(29) | 4.0390(6) | 0.880(3) | -0.4456(26) | 4.0445(6) |
| 6 | 0.875(3) | -0.4429(37) | 4.0363(8) | 0.906(5) | -0.4588(36) | 4.0405(8) |

**Figure Captions**

FIG. 1. (Color online) Sketch of neighboring atoms around an isolated Mn dimer marked by spheres. (a) One-dimensional case studied for $CsMn_xMg_{1-x}Br_3$ (only the Mn and Mg positions are shown). (b) Two-dimensional case studied for $K_2Mn_xZn_{1-x}F_4$ (only the Mn and Zn positions are shown). The full and open triangles denote the ten nearest-neighbor and 14 next-nearest-neighbor positions, respectively. (c) Three-dimensional case studied for $KMn_xZn_{1-x}F_3$ (only the Mn and Zn positions are shown). The triangles denote the 26 nearest-neighbor positions.

FIG. 2. (Color online) Energy level scheme of a Mn(II) dimer for antiferromagnetic exchange coupling (J<0) according to Eq. (1). The higher-lying dimer states with $3 \leq S \leq 5$ are not shown.

FIG. 3. (Color online) Energy spectra of neutrons scattered from $CsMn_{0.10}Mg_{0.90}Br_3$ at T=2 K corresponding to the Mn dimer ground-state transitions $|0,0\rangle \rightarrow |1,\pm 1\rangle$ (a) and $|0,0\rangle \rightarrow |1,0\rangle$ (b). The incoming neutron energy was 2.5 meV. The lines are the result of least-squares fitting procedures as explained in the text. The vertical bars correspond to the probabilities $p_m(x)$ predicted by Eq. (2).

FIG. 4. (Color online) Energy spectra of neutrons scattered from $K_2Mn_xZn_{1-x}F_4$ at T=2 K with incoming neutron energy of 1.55 meV. (a) New data for x=0.02. (b) Data taken from Ref. 2. The lines in (a) and (b) are the result of least-squares fitting procedures as explained in the text. The vertical bars in (b) correspond to the probabilities $p_m(x)$ predicted by Eq. (3).

FIG. 5. (Color online) Energy spectra of neutrons scattered from $KMn_xZn_{1-x}F_3$ at T=2 K with incoming neutron energy of 1.55 meV. (a) New data for x=0.01. (b) Data taken from Ref. 2. The lines in (a) and (b) are the result of least-squares fitting procedures as explained in the text. The vertical bars correspond to the probabilities $p_m(x)$ predicted by Eq. (3).

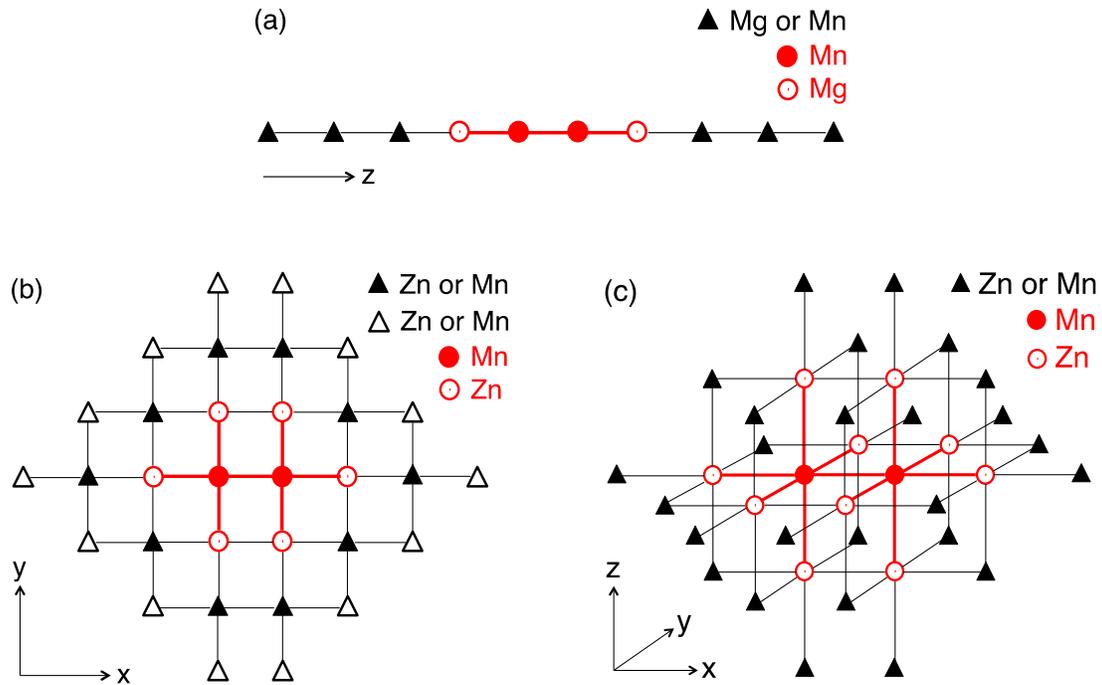

FIG. 1. (Color online) Sketch of neighboring atoms around an isolated Mn dimer marked by spheres. (a) One-dimensional case studied for $CsMn_xMg_{1-x}Br_3$ (only the Mn and Mg positions are shown). (b) Two-dimensional case studied for $K_2Mn_xZn_{1-x}F_4$ (only the Mn and Zn positions are shown). The full and open triangles denote the ten nearest-neighbor and 14 next-nearest-neighbor positions, respectively. (c) Three-dimensional case studied for $KMn_xZn_{1-x}F_3$ (only the Mn and Zn positions are shown). The triangles denote the 26 nearest-neighbor positions.

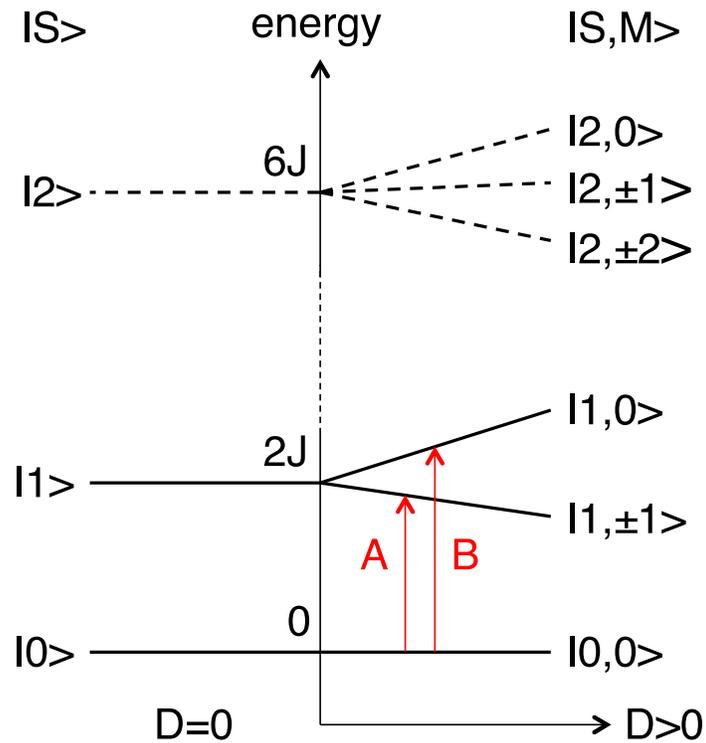

FIG. 2. (Color online) Energy level scheme of a Mn(II) dimer for antiferromagnetic exchange coupling (J<0) according to Eq. (1). The higher-lying dimer states with 3≤S≤5 are not shown.

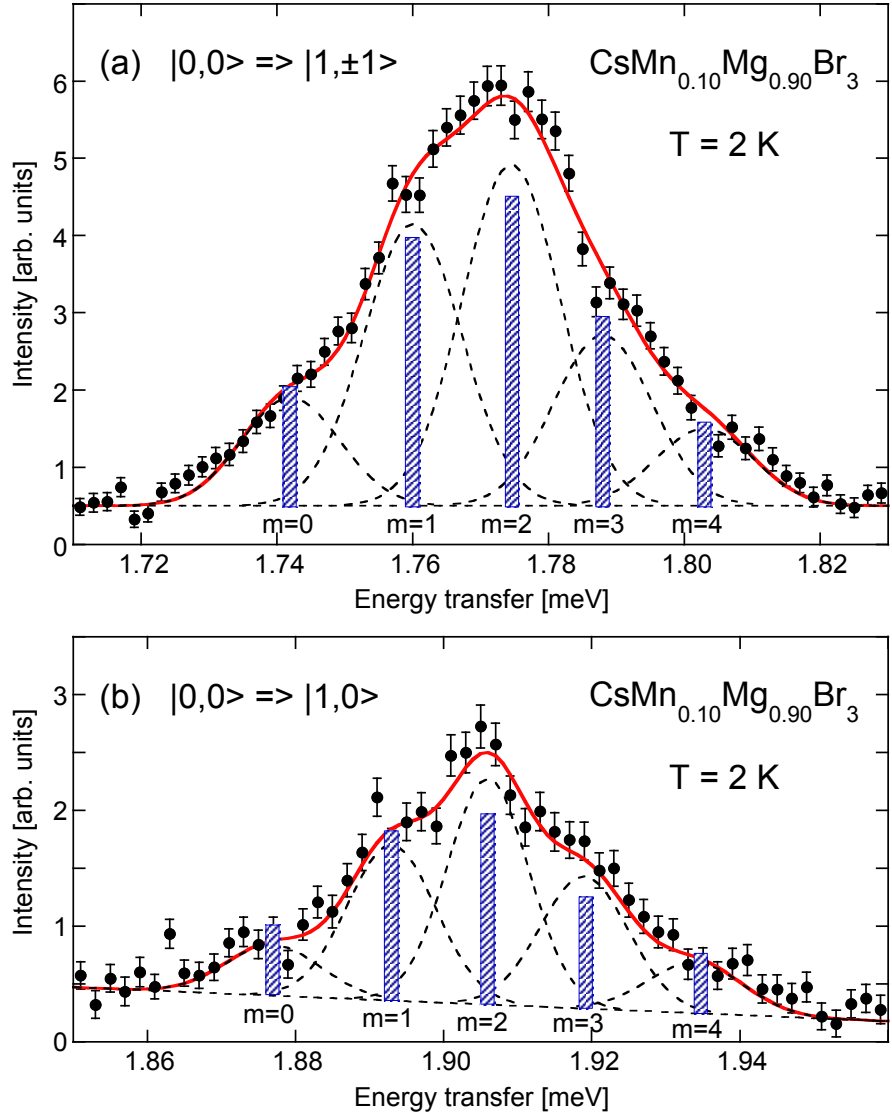

FIG. 3. (Color online) Energy spectra of neutrons scattered from $CsMn_{0.10}Mg_{0.90}Br_3$ at T=2 K corresponding to the Mn dimer ground-state transitions $|0,0\rangle\rightarrow|1,\pm1\rangle$ (a) and $|0,0\rangle\rightarrow|1,0\rangle$ (b). The incoming neutron energy was 2.5 meV. The lines are the result of least-squares fitting procedures as explained in the text. The vertical bars correspond to the probabilities $p_m(x)$ predicted by Eq. (2).

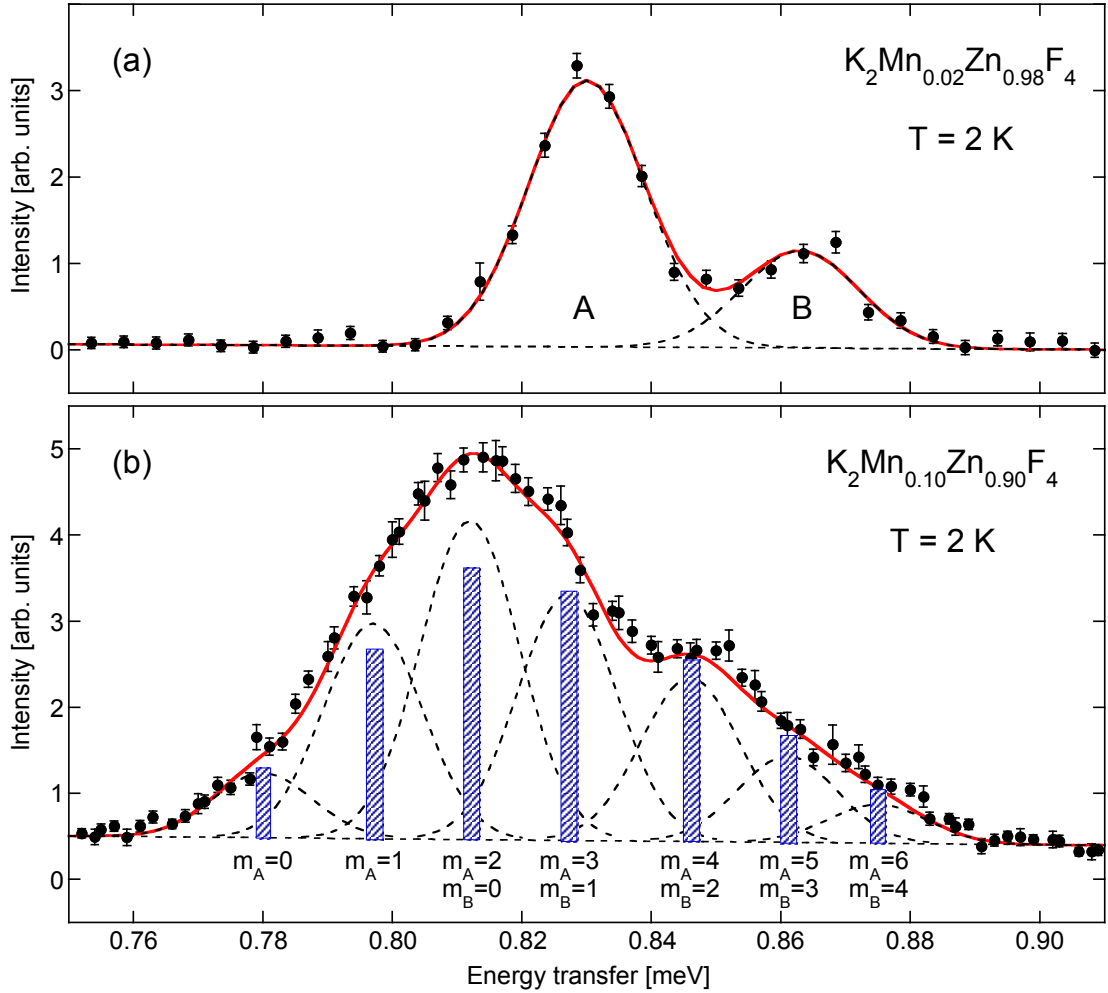

FIG. 4. (Color online) Energy spectra of neutrons scattered from $K_2Mn_xZn_{1-x}F_4$ at T=2 K with incoming neutron energy of 1.55 meV. (a) New data for x=0.02. (b) Data taken from Ref. 2. The lines in (a) and (b) are the result of least-squares fitting procedures as explained in the text. The vertical bars in (b) correspond to the probabilities $p_m(x)$ predicted by Eq. (3).

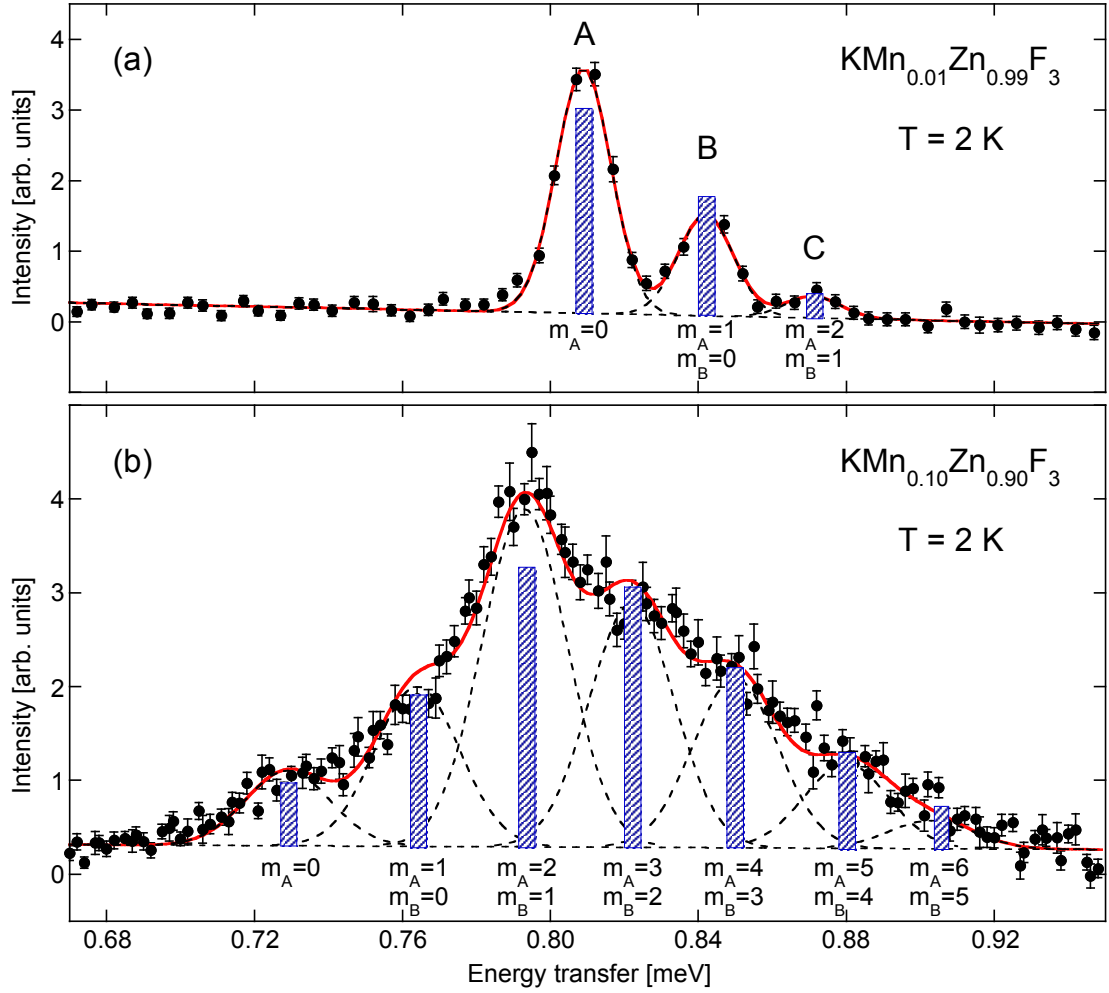

FIG. 5. (Color online) Energy spectra of neutrons scattered from $KMn_xZn_{1-x}F_3$ at T=2 K with incoming neutron energy of 1.55 meV. (a) New data for x=0.01. (b) Data taken from Ref. 2. The lines in (a) and (b) are the result of least-squares fitting procedures as explained in the text. The vertical bars correspond to the probabilities $p_m(x)$ predicted by Eq. (3).